\documentclass[runningheads]{llncs}
\usepackage{graphicx}
\usepackage[inline]{enumitem}
\usepackage{cite}
\usepackage{color}

\begin{document}
\title{An incomplete and biased survey on selected resource management for distributed applications as basis for interoperability of medical devices}
%
%
\author{Imad Eddine Touahria\inst{1,2}}

%
%
\institute{Department of Telematics engineering, Universidad Carlos III de Madrid, Leganes, Madrid, Spain \and
Department of computer science, Ferhat Abbas Setif-1 University, Setif, Algeria  \vspace{5px}
\email{100370038@alumnos.uc3m.es } \email{imad.touahria@univ-setif.dz}\\
}

\maketitle              
\begin{abstract}
\label{sec:abstract}

The popularity and wide spread of IoT technology has brought about a rich hardware infrastructure over which it is possible to run powerful applications that were not previously imagined. Among this infrastructure, there are the medical hardware that is progressively advancing but at a slower pace. Nevertheless, medical devices are more powerful now to run more sophisticated functions and applications and exchange big data with external systems in a secure and safe  fashion.
Towards the design of an architecture for interoperability of medical devices, this paper initially focuses on the background work that is taken by the author for this objective. The paper briefly describes the role of the software in the advances of medical systems and their possibilities for interoperability. It focuses attention on the distribution software layer that is responsible for connectivity, efficiency, and time-sensitivity in the basic operation of medical systems such as exchange of information and commands across devices and systems. The paper analyses a number of previous work on middleware (mostly performed at his research group and also in a broader research community), and pay especial attention to the middleware for web-based systems and how they relate to the development of distributed medical systems.


\end{abstract}
\section{Introduction }
\label{sec:introduction}

As more IoT devices populate everyday life and penetrate on every system and societ, the available hardware infrastructures become progressively richer. The new application paradigms such as \textit{Social Dispersed Computing} introduced in \cite{SocialDispersedComputing} envision the creation of highly complex systems that will provide powerful and intelligent solutions to many of today's problems such as mobility, transport, energy, health, smart living, etc.

The amount of patients in need of continuous monitoring by 2025 will be 1.2 billion \cite{world2003diet} approximately. This indicates the need for providing more sophisticated logic to build intelligent systems with very efficient operation; for this it is needed to find more efficient ways for designing and developing medical devices and their corresponding software that enables continuous monitoring through smart devices (mobile or bedside) that will improve the quality of care and safety.

The integration of newer hardware and software technology in health care systems is an opportunity for decreasing the levels of mortality and  will provide, overall, an improved and smarter patient care.

Health monitoring solutions can be provided on many available devices, from a small medical application deployed on a PDA (Personal Digital Assistant) to measure heart rate to complex systems deployed on a hospital server that ensures monitoring of a patient with critical or chronic diseases and clinician decision support on diagnosis. Medical devices are an example of the new era of smart health care, as these devices offer new possibilities for physicians (on the user side) and provide new opportunities for research on the design and development of health care software that is data-centric and provides efficient  interoperability. 

Critical clinical scenarios are deeply in need of the installation of medical devices especially in cases where patients are located in remote sites and they are in need of diagnosis services in a remote way. The systems that will support these scenarios will have to integrate devices that will be mobile \cite{7336385,8363196,8336776} or fixed close to the monitored patient 
such as the bed or somewhere around her/his home or living site \cite{5070972}. Manufacturers provide numerous alternatives for hardware and software of medical systems with a high degree of variability. These differences may generate some integration problems to build holistic clinical solutions. The  network of the care center (i.e., hospital, etc.) has different computational servers, switches, data recorders and medical devices/equipment; the latter ones have to be integrated in the network as Plug and Play (PnP) devices. Manufacturers have to adhere to standards for medical systems (operation, data exchange, etc.) in order to achieve integrability in the network. 


Improving the assistance given to patients depends greatly on the capacity of medical systems to collect large data amounts in real-time, exchange them, process them, and create new knowledge that could assist the medical decisions. Collecting data from multiple sensors and devices can only be done if \textit{interoperability} is achieved effectively, efficiently, and in a timely manner.

Interoperability is fundamental in the medical domain. At the same time, it is a great challenge given the enormous variety of actors ranging from medical devices to computational units, to human actors. 
As defined in \cite{fdainteroperability}, \textit{medical device interoperability} is the ability to safely, securely, and effectively exchange and use information among one or more devices, products, technologies, or systems. The data that is exchanged can be used for a number of purposes such as display patient information monitoring, storage, interpretation, analysis, and take autonomous action on or control another product. OR.NET \cite{ORNET} and ICE \cite{CIMIT}
are two examples of projects whose objective is to provide a safe and interoperable network of medical devices. This will be later described in this paper.


For safety reasons, medical systems have traditionally been deployed in isolation. However, the proliferation of networking technology and distribution software and middleware has also reached eHealth in such a way that medical systems are becoming more open. To support such interconnection in a \textit{safe} way, the involved actor agreed on the necessity of a common environment to facilitate interoperability and safety. The most accepted one is the Integrated Clinical Environment (ICE), that is a new solution involving a number of stakeholders where the final goal is to realize an interoperable network of medical devices in a \textit{safe} way \cite{8029619}. To fulfill this goal,  manufacturers have to use standards such as ICE (Integrated Clinical Environment) or other medical-device interoperability projects. Communication standards are beneficial for manufacturers, health care providers, and users: According to the west health organization \cite{WEST}, the take up of functional interoperability in medical devices can save up to \$30 billion dollars.

Having said the above, this paper focuses on the impact of the software technology on the development of intelligent health solutions and their associated medical systems. By making a biased survey centered in the work of some researchers, the paper exposes selected contributions of a limited part of the research community that target at developing distribution software; this is an exercise to analyze these specific contributions that feed the author's background and  will take them as basis for producing a new enhanced solution tailored to the needs of medical device interoperability.

\subsection{Paper structure}
\label{sec:structure}

The paper is structured as follows.  
Section \ref{sec:rolemw} describes the role of the distribution software for developing distributed medical applications, emphasizing the web technology that facilitates interoperability via standard protocols.
Section \ref{sec:group} describes state of the art on middleware design and distribution, focusing on the work of the author's group that serves as his reference for improving middleware design for medical systems;  also, other technologies are listed that are relevant for his current work.
Section \ref{sec:standards}  describes illustrates some examples of frameworks for medical devices coordination and interoperability projects.
Section \ref{sec:conclusion} 
 presents the conclusions and the future work that will be part of the author's thesis.

\section{Distribution software design for medical systems}
\label{sec:rolemw}

Medical systems experiment the same difficulties of the rest of systems as far as its interoperability is concerned. Connectivity problems arise due to incompatibility of the software level (e.g., operating system, networking protocols, format of exchanged data, etc.). This problem has been well identified by the community that develops and designs medical systems and a number of projects have appeared that are studing the way of solving interoperability problems and providing safe solutions at the same time. The distribution software typically takes care of these type of technical issues, so the middleware community has a lot to say in this picture.
Solutions will have to consider ICE, that is the most popular and widely adopted standard for medical device interoperability. Most of its implementations are based on publish-subscribe middleware such as the Data Distribution System \cite{DDS}. Nevertheless, there is a long road towards obtaining proper solutions for efficient communication and interoperation in a timely, flexible, and reconfigurable manner.

On the author's road to provide one of this solutions, this paper performs a study of a reduced set of solutions in the research to design and develop real-time middleware. The paper analyzes the research on distribution software of his reference distributed real-time systems group that will be used as the background to the research on this field introduced there. This work has the intention of clearly exposing some selected problems and solutions to real-time middleware design for general distributed applications and also for cyber-physical systems. The analyzed work starts with techiques for resource management that are focused at operating system level, the design of software engineering solutions for middleware, service oriented techniques, real-time reconfiguration, modeling of cyber-physical systems and real-time online verification, and hardware acceleration embedded into the middleware logic. Taking this as a solid baseline, the author will extend these works to suit the needs of medical systems. 

\subsection{Middleware definition}
\label{sec:middefi}
The term \textit{"middleware"} has been used under different contexts. At first, it was described as a layer of software located between platforms and applications in distributed systems \cite{Bernstein1996MiddlewareAM};  it is widely defined as an software layer that connects two or more heterogeneous applications, systems, or structures \cite{Bishop03asurvey} which, by the end, provides an interface to transfer data and commands among the different stakeholders. In the medical domain, HL7 is one of the most important middleware technologies (referred to in sections \ref{sec:networking} and \ref{sec:interoperability2}).

Middleware is a reusable software layer that renders standard interfaces and protocols to frequently encountered problems like heterogeneity, interoperability, scalability, fault-tolerance, security, resource-sharing \cite{5635187}. The middleware technology aim to protect the application layer from problems generated by the lower layers and that are devices heterogeneity, data encoding, security encryption algorithms. Also, middleware constitutes a way to achieve \textit{interoperability}, and this is done by handling the heterogeneity of computer architectures, operating systems, programming languages, and networking protocols to facilitate application development and management \cite{Geihs:2001:MCA:619064.621730}.

\subsection{Middleware technologies}

\subsubsection{RMI.} RMI (Remote Method Invocation) \cite{RMI} is a JAVA API (Application Programming Interface),and as it name indicate its used to call remote methods, [8] defines RMI as a technology to create communication between Java-based applications,  where Java objects can beinvoked on the same machine or on different machines deployed on different networks, itsupports object polymorphism, its easy to write, secure and mobile, and supports dis-tributed garbage collection.  RMI is based on TCP/IP technology and was proposed likea version for the method called RPC (Remote Procedure Call).  The two machines thatare exchanging data with RMI technology have to run a JVM (Java Virtual Machine),RMI is object oriented and implement the client/server technology.
\subsubsection{CORBA}. CORBA (Common Object Request Broker Architecture) \cite{CORBA} is an OMG standard and is one of the first developed middlewares, it allows different softwares or applications developed with different programming languages and present in different locations to communicate together via an interface broker. The core concept in CORBA is Object Request Broker (ORB), ORB allows a client application to request services from a server application without knowing the details of the application server or the configuration of the network where it's deployed. CORBA uses Interface Description Language (IDL) to define the interfaces of objects which facilitates the communication between different actors.
\subsubsection{iLand.} iLand \cite{6198329} is an open source middleware that has been applied in industrial prototypes, including medical systems. It  follows the classical principles of a layered middleware design; though its architecture is independent of the underlying communication network protocol, the reference implementation of iLand uses a DDS backbone. iLand includes a number of enhanced functions to support dynamically re-configurable applications based on services: light-weight services in the real-time version and web services in the soft and best effort version with QoS guarantees.
\subsubsection{DDS.} DDS stands for Data Distribution Service \cite{DDS}, it is a middleware protocol and API standard for data-centric connectivity provided by OMG (Object Management Group) standard. DDS serves as middleware architecture for a publish/subscribe messaging pattern and integrates the components of a system together, providing low-latency data connectivity, extreme reliability, and a scalable architecture that business and mission-critical Internet of Things (IoT) applications need. DDS is \textit{data centric}, which is ideal for the Internet of Things. Data centricity means that DDS knows what data it stores and controls how to share that data.


\subsection{Middleware position in distributed medical solutions}
\label{sec:midpos}

The development of distributed applications has been greatly enhanced by middleware technology, one important realization of this last in medical systems is the automatic handling of patient records. Current practice often uses health information systems (HIS) and electronic health record (EHR) in an informal manner with adhoc protocols and interoperability solutions in order to develop clinical systems \cite{marisolmedCP20162}.

Middleware technology has an important role in medical systems in what follows we mention some of those roles:
\begin{list}{$\bullet$}{\leftmargin=1em \itemindent=0em}
\item Improves data transfer mechanisms between medical devices and the computational servers where applications are deployed.
\item Defines a secure layer that defines algorithms and encryption methods and therefore enhance data security.
\item The programmer is more concentrated on the programming details because the hardware details are abstracted by the middleware. 
\item Achieves interoperability by connecting medical devices, computational servers, applications and data storage solutions.
\item Patient data must be private, middleware can be the intermediate that ensures a certain level of privacy and especially when data is transferred into third party stakeholders. 
\item Improves the portability of applications \cite{4519606}, where an application can be executed under many plateforms. 
\end{list}

Over the last years, middleware has proven successful to address the ever increasing complexity of distributed systems in a reusable way \cite{Issarny:2007:PFM:1253532.1254722}, thus, middlewares are taking an important place in complex systems design like medical applications that controls medical devices that have a direct impact on patient safety or a medical system that controls drug injection in patient body.

Middleware is characterized by abstracting the low level details of the communication protocols and the hardware characteristics of devices to programmers \cite{sensorsice}. The distributed nature of medical applications requires the use of technologies such as middlewares that provides a bridge between layers or levels, the hardware layer can be connected to the application layer via a \textit{middleware} and the protocols level can be connected to the security level via a \textit{"middleware"}, so the use of the middleware depend on the context of it application. By abstracting the low layers problems and concentrating on the application layer the programmer can generate better solutions. According to \cite{4809544}, middleware technology can be classified into tow types and this depends on the level of use, \textit{low level}  middleware exists between sensor nodes and medical devices and \textit{high level} middleware connects diffrent application and tra,sfer data between them.
\label{sec:midexa}

\section{Baseline}
\label{sec:group}

{\color{black}
This section analyses the state of the art work on which the author will build on. Here, it is described the work of the author's group related to the design of resource management strategies for middleware in order to build time-sensitive and efficient systems, as well as real-time middleware design for cyber-physical systems and medical systems.
}

\vspace{0.2cm}

\noindent \textbf{Resource management and components for real-time}. This section presents techniques and algorithms to manage the computational resources; it constitutes the lowest abstraction level related to efficient resource assignment and enforcement techniques to avoid interference in the execution among several application tasks.

Controlling the execution at thread or task level to make an efficient resource usage is essential in distributed applications. There are a number of contributions on the architecture of real-time middleware from a real-time perspective. One of the first ones was named Hola-QoS described in \cite{HolaQOS} and \cite{the01}; using real-time scheduling for distributed actions \cite{ares}; real-time quality of service management \cite{ambient}; mode changing policies for timely execution \cite{modechanges,dualBand}; agents modeling real-time properties \cite{mata}; identification of Linux kernel properties for improving locking \cite{peteradaeurope}; architecting open source projects for Linux \cite{peterspe}; analysis of temporal behavior at bus level within a multiprocessor \cite{groba}; reconfiguration scheduling \cite{aina}.

Other contributions for service oriented systems have been \cite{soca}; or component-based modeling over QoS networking \cite{demiguel}; have progressively shed light over how to handle execution in systems using more abstract modeling like encapsulation through services and separating application from networking responsibilities.

On the distribution software side, a number of works have been published to fine tune the resource management policies inside the distribution software. On \cite{canojspe} and \cite{canocorr}, a solution to build real-time component replacement for OSGi systems is provided. The work \cite{iadis} presents a higher abstraction to separate concerns in managing thread-level resources flexibly. Also, \cite{qosbidimensional} provides a higher abstraction resource management policy based on a quality of service specification inside a middleware to flexibly reconfigure applications based on services. Based on this, \cite{RTServiceComposition} provides a survey on real-time service composition in distributed systems and proposes a new solution to achieve real-time composition, further enhanced by \cite{mgvj27} laying down different alternatives for reconfiguration; and \cite{c29} describes two alternatives for application reconfiguration built inside the middleware one at task level and the other at service level. Other approaches describe garbage collection techniques inside the middleware such as \cite{j22} that support real-time execution even during system maintenance time.

Other works provide higher level designs of specific applications based on real-time middleware such as \cite{j16} that uses the Data Distribution Software for remote control.

As part as the new virtualization paradigms that provide new execution infrastructures where applications can coexist, we find \cite{ddscsi,c35} that studies the performance and \cite{j26} that provides predictable cloud computing.

\vspace{0.2cm}

\noindent \textbf{Cyber-physical systems and IoT spheres}. Cyber-physical systems and IoT are highly related because both refer to the monitoring and actuation over physical objects. In the case of medical systems, the physical objects are the patients and the sensors and actuators that monitor them are medical devices (that are very close to IoT devices). Medical devices are part of a medical cyber-physical system (MCPS) that monitor the physical conditions of patients and actuates on them or help in deciding how a human physician will actuate on the patients, providing advice for a recommendation system or a decision support system. Cyber-physical systems have to employ  rigorous design techniques because they are critical systems; for this, they have to use formal methods to verify that the system properties are kept at all times. 

CPS have to be systems capable of evolving and should, therefore, be flexible. This is so because CPS in the medical domain target the monitoring of humans; humans have mostly an unpredictable behavior and this may raise unstable situations and unexpected events can be triggered. This means that these systems cannot be designed offline fully as not all the situations are known at design time. CPS have to  support evolution and flexibility that are important challenges. It is extremely challenging to integrate the uncertainty raised by the unknown monitored conditions of a patient, etc., and the real-time reaction that is needed, as all the possible situations that could happen when the system is in operation can not be known a priori. In \cite{JSSPoliMi}, a formal design is described based on Petri nets to model systems that can evolve; this technique was also explored in \cite{Compsac2014}. A different formal method approach is applied in \cite{Bersani18} and \cite{BersaniHASE18}. Also, \cite{c38} inserts online verification mechanisms based on Petri nets inside a distribution middleware.

In the above works, the focus is placed on the design of the software component interactions relative to their timing properties and other behavioral parameters that are modeled. However, the communication across the above components is a key aspect that must be analyzed to achieve communication and interaction infrastructures that support timely interaction and variable conditions such as load peaks or coexistence of components with heterogeneous resource usage patterns.
For this purpose, there are also a number of contributions for the design of distribution middleware for CPS as middleware is the key software layer that is capable of abstracting distribution and interaction, masking situations where a node can receive a peak of requests from other nodes; the systems must be resilient to these and other situations, and they have to continue to work at all times. The design of adaptive middleware is provided in OmaCy architecture \cite{omacy}. In \cite{FGCSreview} and \cite{JSAReiser}, an analysis of this problem is outlined. In \cite{CSIcpsmiddleware}, the design of a scheme for attending simultaneous requests is provided. In \cite{RPiDDS}, a model for integrating the Data Distribution Software with single board computers and Raspberry Pi is provided; this is further reworked in \cite{DDSFace} for a different domain such as avionics. Also, there have been a number of dedicated research contributions to building real-time facilities in middleware such as
\cite{jsareaction2012, JSAReiser}, among others; or building abstractions for utilization of multiple interaction paradigms such as \cite{adaeurope11} or \cite{adaeurope12}.

\vspace{0.2cm}

\noindent \textbf{Medical systems}. It is fundamental to analyze the position of middleware within medical systems in order to develop safe execution solutions that also provide timely operation. Failure to meet these requirements may yield to hazards to lives. Service-oriented architectures such as iLAND \cite{ilandTII} (that has been well proven in a number of critical domains) have been integrated with ICE in \cite{sensorsice}. The reconfiguration capacity \cite{isorcyork} and timely communication of iLAND have been proposed to be the core interoperation backbone for ICE. A number of studies for profiling the actual performance of communication middleware such as \cite{uc3muma} has been particularized for medical systems in a number of works such as \cite{sigbediland} for the Internet Communications Engine and \cite{sigbedamqp} for AMQP. Moreover, a number of improvements to their execution by making the middleware aware of the underlying hardware structure have been undertaken in \cite{sac2017} and the benefits of this acceleration in medical systems for remote patient monitoring has been exemplified in \cite{uc3mupv} for eHealth services and in \cite{ades} for audio surveillance.

It is important to bare in mind that designing medical systems requires detailed design and validation of properties as they are critical systems. There are different design and development frameworks for critical systems that have to be applied considering not only the functional properties of the application-level logic, but also the non-functional properties of the whole software stack. Some frameworks exist like \cite{infsof, c44} that support web-based monitoring of critical software projects like the development of medical systems, including their set of libraries like the distribution middleware.

\vspace{0.2cm}

\section{Key definition of \textit{Medical Devices}}

According to the WHO (World Health Organization) definition of ``medical device'' \cite{WHODEVICE}, there are a number of possible hardware systems containing software and hardware parts that sample patient data through reading of vital signs, use networking protocols to exchange data or share it, and have different posibilities on mobility.



The development of medical devices undergoes strict regulations that are established by the FDA (Food and Drug Administration) \cite{FDA}. The role of FDA has faced criticism on its lack of innovation due to the strict regulations; this is the case of X-Ray machines innovation critics \cite{ekelman1988technological}. On the other side of the Atlantic, the European Union adopted the MDD (Medical Device Directive) as the governing set of standards and regulations for medical-devices manufacturer. Some of these directives are the Medical Device Directive (MDD 93/42/EEC), the In Vitro Diagnostic Medical Device Directive (IVDMDD98/79/EC) and the Active Implantable Medical Device Directive (AIMDD 90/385/EEC) \cite{french2012medical}. Overall, in different countries, regulation have gone through different paths; however, those countries most involved in medical device regulation established the Global
Harmonization Task Force (GHTF) and, after that, the International Medical Device Regulators Forum (IMDRF) \cite{IMDRF} appeared with the goal of enforcing a faster medical device regulatory harmonization and convergence at international level in what concerns safety, performance and quality of medical devices.

According to the EU Borderline Manual \cite{manualborder}, the following types of software should generally be classified as medical devices:

\begin{itemize}
    \item picture archiving and communication systems;
    \item mobile apps for processing ECGs;
    \item software for delivery and management of cognitive remediation and rehabilitation programs;
    \item software for information management and patient monitoring; and
    \item mobile apps for the assessment of moles (e.g. making a recommendation about any changes).
\end{itemize}

Also, the EU provides recommendations on types of software that should generally not be considered as medical devices. These are  mobile apps for communication between patient and caregivers while giving birth or for viewing the anatomy of the human body; software for interpretation of particular guidelines or mobile apps for managing pictures of moles like recording updates and changes over time.

As indicated by  the World Health Organization \cite{WHODEVICE}, software can be considered as a medical device. Moreover, IMDRF \cite{IMDRF} defined the concept of \textit{``Software as a Medical Device''} that is a software intended for use in different devices as medical functions that perform them without being part of a hardware medical device. This follows the classical role of the software engineering vision and software as a service paradigm.  It should be noticed that this study considers that a software is a medical device when the functional properties of the software are enough to handle a clinical situation in which the software service is used as a whole medical device.

In the context of this paper, software is a key part, and middleware is a part of software stacks. As the importance of software increases in the health domain \cite{6334808}, the importance of middleware also raises. Given the cyber-physical systems relation to medical devices, also the medical software has vital safety requirements that force it to adhere to strict parameters concerning data accuracy, integrity, security, and verification. This means that its design should follow strict rigorous validation and verification techniques just as CPS do.

\subsection{Analysis of medical devices characteristics}

\subsubsection{Safety.}

\textit{Safety} \cite{7814546} is the most important parameter in the development of medical systems. For this, their design must comply to standards like: European MDD \cite{199342EC, 200747EC}, ISO 13485 \cite{134852012}, IEC 62366 \cite{623662007}, ISO 14971 \cite{149712012} and others.

The high variability (e.g., displays, sensors, actuators, communication capacities, and even materials) across different devices challenges their design. 

To overcome the communication challenges, medical devices  are progressively provided with standard input/output ports, leaving behind the propietary data output ports that would highly vary across suppliers of medical devices. The most common ports are RS 232 port (DB-9, DB-15, and DB-25),
RJ 45, wireless LAN, Bluetooth, USB, or some proprietary data connection systems
developed by suppliers for using data by their own IT systems. The following are the most common hardware connections used by suppliers to input data into the device: PS/2 (for
supporting  keyboard or a mouse inputs),
USB, RS 232, and digital data input \cite{8029619}.

\begin{figure}[h!]
\centering
\centerline{\includegraphics[scale=0.5]{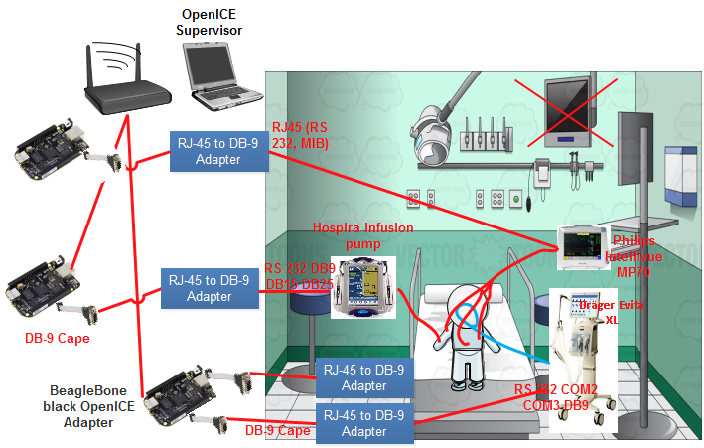}}
\caption{Hardware configuration of medical devices according to ICE architecture}
\label{fig:hardware}
\end{figure}

\subsubsection{Timing requirements.}
\label{sec:time}

Medical systems are time-sensitive. It is the case that different systems have different requirements for timing behavior, ranging from QoS (quality of service) and best effort to hard real-time guarantees. Therefore, medical devices should support different levels of real-time guarantees in the collection of data, the processing of data, visualization of medical data and alarms, and actuation on the patient or system. The respect of timing requirements by a medical devices is related to the ability of it software to treat and transfer data with consideration to the quality of it hardware components and the good installation of electrodes and cables.

The timing requirements of a medical device include the execution time of an operation (C), deadline (D), and the period (T) of an operation \cite{8029619}, these timing requirements are defined ate design time (requirements collection phase) and verified at run time (system operation). Manufacturers can only specify and guarantee the
individual devices’ standalone real time properties which can be distributed via ISO/IEEE 11073 medical device description format \cite{7844043}, but, when connected in a network or to other devices the real timing measurements can be get after executing the system.

\subsubsection{Electromagnetic compatibility.}
\label{sec:Electromagnetic}
In some clinical cases, more than one medical device can be integrated into the same OR (Operating Room), also, many medical devices can be used for different patients into a small clinical spot. Electromagnetic fields (EMF) may cause problems on electronic devices \cite{doi:10.1080/10803548.2009.11076785} and generate safety hazards in the medical devices, the level of those hazards increases when the affected MD has an important role in the treatment process, such as: implantable infusion pumps and cardiac peacemakers. An example of a hazard generated by radio-frequencies is an overdose of insulin from infusion
pumps exposed to EMF from mobile phones or RFID (radio frequency identification devices, usually emitting RF EMF
of 0.8–2.4 GHz) \cite{karpowicz2013assessment}.

The radio-frequencies generated by a medical device or another source (infrared, microwave or ultrasound therapy) can affect the patient body and the body of others-the physiotherapist or bystanders, if they find themselves within the EMF \cite{Gryz2014EnvironmentalIO}, thus, medical devices must be designed to operate in specified intervals of radio frequencies where their behavior will not be affected by the differences of frequencies. The impact of EMF can generate wrong results of medical devices, the deterioration of the hardware components of another MD, and also can cause cancer for clinicians that are exposed for long periods to EMF.

\subsubsection{Data accuracy.}
\label{sec:accuracy}
The final goal of a medical device is to monitor patient safety, inject drugs into patient body in some cases, intervenes in surgical operations, and provide data that is accurate for ulterior use by clinicians and staticians. The big flow of data in healthcare applications generates a need to have well-defined description of rich and structured data required to represent the variety of data used in clinical environment \cite{weiningerintegrated} and thereby data that is \textit{correct}. To have accurate data the clinician responsible of the good use of the device must insure the good installation of electrodes and cables on the patient body, the appropriate environmental conditions to the MD and the absence of interoperability issues.
\label{sec:Vulnerabilities}

\section{Interoperability alternatives}
\label{sec:standards}

HIMSS (Healthcare Information and Management Systems Society) \cite{himms}
defines \textit{interoperability} in healthcare at three levels:
\begin{itemize}
    \item Data exchange between two different information technology system. This includes the ability to exchange the data, to receive it and to interpret the data. This is a foundational interoperability level.
    \item Structure or format of the exchanged data. This refers to the standards for message formats. It implies that the data content and meaning, as well as its operational purpose, is preserved and unaltered. 
    \item Semantic. This level leverages the two levels above and the coding of the data (that includes its vocabulary) in order to allow the receiver systems to interpret the data. 
\end{itemize}

All these levels are common to general purpose middleware applied to the specific domain of health care.

\subsection{Data exchange standards}
\label{sec:networking}

Communication and data exchange between medical devices is a basic support for \textit{interoperability}. The basic data exchanged is the EHR. As such, there were a number of EHR incentive programs for building efficient data management in this respect. Two health IT standars competed years ago for this purpose: 

\noindent \textbf{CDA}. This stands for Clinical Document Architecture that was backed up by the Health Level Seven International (HL7).

\noindent \textbf{CCR}. This stands for Continuity of Care Record that was empowered by the American Society for Testing and Materials (ASTM).

There were a number of proposals for harmonizing the two; however, CDA won over CCR in a first battle. Nowaday, it exists \textbf{C-CDA} that stands for Consolidated CDA format that is e by the National Coordinator for Health Information Technology (ONC).

Some other  alternatives are shown below.

\noindent \textbf{DICOM} (Digital Imaging and Communications in Medicine) is a standard developed by the National Electrical Manufacturers
Association (NEMA) to store, retrieve, and transfer medical digital information exchange, basically imaging data from imaging equipment, from various medical devices such as scanners, printers, network hardware etc \cite{7988050}.

\noindent \textbf{HL7 version 2} 
is a standard designed for messaging in a centralized or distributed environment. Also, it supports the proposition of interfaces for communicating to third parties that do not adhere to data exchange standards. HL 7 standard is one of the most popular ones; precisely, 95\% of hospitals, 95\% of medical related equipment and information systems in the whole America use it. It is also in use in Germany, Japan and other developed countries \cite{5974909}. 

\noindent \textbf{HL7 version 3} is a newer version of HL7 version 2 that uses the eXtensible Markup Language (XML) as a powerful tool in the web to transfer data. Also, version 3 integrates web services and the Web Services Description Language (WSDL) \cite{5687707}. Version 3 of HL7 promotes semantic interoperability defined a more explicit methodology for the development of messages \cite{7377675}.

\subsection{Health specific seamless-communication standards}
\label{sec:interoperability2}

Following, a number of protocols that provide seamless interoperability in medical systems is provided.

\noindent \textbf{HL7}. This standard is again listed here as its main goal is to provide seamless integration across a network of medical devices and also in a way that is fully secure. HL7 defines data transfer formats for interoperability across medical devices and HIS (Health Information Systems). 

\noindent \textbf{ISO/IEEE 11073} designates a set of standards for plug and play interoperability of medical devices. ISO/IEEE 11073 defines a common framework for the establishment of a unified data structure model \cite{7494019}. ISO/IEEE 11073 (X73PHD)\cite{11073,staff2010iso} objective is to standardize Personal Health Devices (PHDs) and allow semantic interoperability of medical devices by defining the structure of data and the protocol for information delivery between individual
medical devices (such as Glucose meter, Weight scale, Blood pressure
monitor, etc.) and the manager (computer, smart phone, set top box, etc.), which collects and manages the information from the individual medical devices \cite{6152915}. 

\noindent \textbf{FHIR} (Fast Healthcare Interoperability Resources) of HL7. It is an environment and framework for EHR exchange that integrates both the market needs and the more up-to-date technologies \cite{8031128}. FHIR targets REST (Representational State Transfer) architectural style as presented by Fielding \cite{Fielding:2000:ASD:932295}. For this purpose, it models the actors of healthcare scenarios as resources: medical software, clinicians modeling, medical devices, medications, or IT structures, among others. FHIR is the most recent standard of the series HL7 v2, HL7 v3, CDA developed by HL7 \cite{6627810}. 
\subsection{Web-services based interoperability}

\noindent \textbf{DPWS} (Devices Profile for Web Services) uses SOA (Service Oriented Architecture) for providing interoperable, cross-platform, cross-domain, and network-agnostic access to devices and their services \cite{7325198}. DPWS is used for embedded devices with limited resources by enabling Web services using IoT applications. DPWS requires WSDL (Web Service Description Language) and SOAP (Simple Object Access Protocol) to communicate the device services, but it does not need a registry like UDDI (Universal Description, Discovery and Integration) for services discovery. DPWS aims to achieve interoperability by using the loosely coupled concept of Web services over the MD operation and data encryption.

\noindent \textbf{MDPWS} (Medical Devices Profile for Web Services) is a part of IEEE 11073-20702 series of standards and uses the principles of DPWS but for medical devices interoperability domain with some modifications like the restricted security mechanisms of MDPWS comparing with DPWS, e.g. the usage of client authentication with HTTP authentication is withdrawn in favor of using X.509.v3 certificates \cite{7390446}. MDPWS uses the principles of DPWS with respect to the high acuity patient environment and the complexity of medical devices.

\subsection{Software frameworks for medical systems}

\subsubsection{OSCP}
\label{sec:OSCP}

OSCP (Open Surgical Communication Protocol) \cite{oscpde} is based on the data transmission technology Medical Device Profile for Web Services (MDPWS, standardized as IEEE 11073-20702) and the Domain Information and Service Model (standardized as IEEE 11073-10207). As mentioned before, MDPWS is based on a SOA architecture and it allows devices to detect and find other medical devices in a local network using WS-Discovery.

The description of devices is based on formal notations and the creation of device description templates. As it is based on formalisms, it supports using different logics to verify correctness and properties of the systems and the devices such as LTL (Linear Temporal Logic, Smart Assertion Logic for Temporal Logic (SALT), regular LTL (RLTL), etc.

\subsubsection{ICE}
\label{sec:ICE}

(Integrated Clinical Environment) \cite{CIMIT} architecture was defined in 2009 in ASTM (American Society for Testing and Materials) F2761 standard. ICE was designed for bridging the gap defined by the high heterogeneity of medical devices. ICE aims at providing a set of specifications and architecture that implements a plug and play atmosphere to create networks of medical devices and to create a communication gateway between them, where messages and commands are exchanged appropriately. 
To survive to the next generation of medical applications and systems, manufacturers will have to adhere to the ICE specifications. One of the most important design decisions of ICE is that medical devices have a network output port and must produce data that can be managed through ICE interfaces. The following are the main ICE objectives:

\begin{list}{$\bullet$}{\leftmargin=1em \itemindent=0em}
\item Improve patient safety by coordinating medical devices actions and avoid incorrect medical decisions generated by a faulty device operation.
\item Ensure support for clinicians in their monitoring and treatment operation, where clinical aid information is generated by a set of workflows implemented in the ICE framework logic.
\item Create a flexible communication bus between medical devices, servers running medical applications, and the clinicians.
\item Implement an interoperable network of medical devices and computational servers where data and messages are exchanged in real time.
\item Define standards for the hardware and software characteristics or dimensions of medical devices that will be used by manufacturers to produce  medical devices that comply with ICE.
\end{list}

 In ICE-based systems \textit{safety} is the ability to implement interoperability between heterogeneous medical devices in a single high acuity patient environment where communication is done via software or hardware interfaces. ICE aims to improve patient \textit{safety} by elaborating and deploying interoperability of the medical devices, thus, creating an interoperate communication bus between heterogeneous medical devices where messages and commands are exchanged.

\begin{figure}[h!]
\centering
\centerline{\includegraphics[scale=0.9]{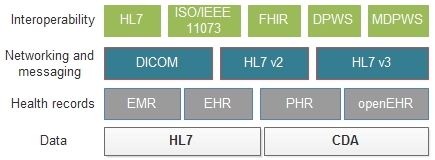}}
\caption{Hardware configuration of medical devices according to ICE architecture}
\label{fig:hardware}
\end{figure}

\subsubsection{OR.NET.}
\label{sec:OR.NET}
OR.NET is a solution developed by German academics and industrials for medical devices integration and medical systems interoperability in the operating room and it surroundings. The objective of OR.NET is to develop basic concepts for the secure dynamic networking of computer-controlled medical devices in the operating room and clinic  \cite{ORNET}.  In the end, these concepts are evaluated and transformed to standards. OR.NET aims to create a service-oriented architecture for the safe and secure dynamic interconnection of medical devices in the OR context \cite{7318708}. 




Besides OSCP, 
OR.NET also allows the use of different communication
protocols (e.g. DICOM and HL7 Version 2). OSCP explicitly
does not try to replace these widespread protocols. Instead, dedicated
gateways are specified that enable the operation of the
DICOM and HL7 protocols despite the separation of OR network
and the hospital network.


\section{Conclusion and future works}
\label{sec:conclusion}

This paper has analyzed the baseline for design of middleware that applies both to general distributed applications and it is later taken to constitute a take-off platform for desiging the needed adjustments for medical systems. The analysis is confined to the research group of reference for the author, comprising the fields that are part of embedded research curricula \cite{j1} such as  resource management techniques, middleware design, cyber-physical systems, and software engineering techniques for functional design considering non-functional properties. Also, technologies related to medical devices are studied in details from data retrieving to processing and final decisions for clinicians. Requirements of medical devices good operation are presented and interoprability as the core concept in medical devices communication is detailed with examples. 

This work is the base for future researches aiming to study and define the position of middleware technology in medical systems and especially in medical devices communications, here ICE is the objective of development. An architecture for ICE-based systems development is under study, this work is the outline for medical devices communications and data exchange through real implementations in clinical environment.

\label{Bibliography}
\bibliographystyle{plain}
\bibliography{References}
\end{document}